\documentclass[10pt]{iopart}
\usepackage{graphicx,bm,amssymb}
\eqnobysec

\newcommand{\beq}{\begin{equation}}
\newcommand{\eeq}{\end{equation}}
\newcommand{\beqa}{\begin{eqnarray}}
\newcommand{\eeqa}{\end{eqnarray}}

\newcommand{\abs}[1]{{\left\vert#1\right\vert}}
\newcommand{\dd}{{\rm d}}
\newcommand{\ii}{{\rm i}}
\newcommand{\mean}[1]{{\langle#1\rangle}}
\newcommand{\prob}{\mathbb{P}}
\newcommand{\blanc}{{\hskip 4pt}}
\newcommand{\E}{{\cal E}}
\newcommand{\1}{{\bm 1}}

\begin{document}

\title{The first positive position of a lattice random walk}

\author{Claude Godr\`eche and Jean-Marc Luck}

\address{Universit\'e Paris-Saclay, CNRS, CEA, Institut de Physique Th\'eorique,
91191~Gif-sur-Yvette, France}

\begin{abstract}
The distribution of the first positive position reached by a random walker starting at the origin
is central to the analysis of extremes and records in one-dimensional random walks.
In this work, we present a detailed and self-contained analytical study of this distribution
for symmetric finite-range lattice walks,
whose steps are drawn from a distribution supported on finitely many integers.
\end{abstract}

\eads{\mailto{claude.godreche@ipht.fr},\mailto{jean-marc.luck@ipht.fr}}

\maketitle

\section{Introduction}
\label{intro}

We consider a random walker on the one-dimensional lattice,
whose integer position at discrete time $n$ is given by
\beq
x_n=x_{n-1}+\eta_n.
\label{xdef}
\eeq
The walker is launched from the origin ($x_0=0$).
The steps $\eta_n$ are independent, identically distributed (iid) integer random variables drawn from an even distribution
supported on finitely many integers:
\beq
\prob(\eta_n=k)=\rho_k\qquad(k=-K,\dots,K),
\label{rhodef}
\eeq
with $\rho_k=\rho_{-k}$.
The integer $K\ge1$ is called the range of the step distribution.
We assume that $\rho_1$ is non-zero,
i.e., that steps with unit length are allowed.
This hypothesis will allow us to discard non-generic behavior.
The probability $\rho_0$ for the walker to stop at every time step may either be zero or not.

The main subject of this paper is the distribution of the first positive position of the walker, denoted by $H$.
This distribution is fundamental for understanding the statistics of successive record positions, or `ladder heights',
in one-dimensional random walks\footnote{See~\cite{revue} for a review on records for random walks and L\'evy flights.}.
The present work is a direct continuation of our recent comprehensive investigation
of the same quantity in continuous symmetric random walks~\cite{I}.
The latter provides a detailed account of the subject, including its history,
and sets key observables in perspective with one another.
Most of the bibliographic references cited therein will not be repeated here.

In the present situation, namely symmetric finite-range lattice walks,
the Wiener-Hopf equations governing the problem can be solved analytically,
without having to resort to the general formalism.
The essence of the method is to factorise the relevant analytic functions over their complex zeros and poles.
This approach was put forward long ago by Wick~\cite{wick} and Chandrasekhar~\cite{chandra}
for Wiener-Hopf equations of the kind encountered in continuous random walks (see~\cite{chandrabook} for a review).
In that setting, the method works whenever the step distribution is a finite superposition of decaying exponentials,
so that its Laplace transform is a rational function.
The possibility of applying a similar approach for finite-range lattice walks
is mentioned in Feller's book~\cite[Ch.~XII]{feller2}.
The method has been rediscovered and given the name `algebraic trick'
in a study of the crossings of a biased simple random walker with a ballistic obstacle
moving at rational velocity~\cite{sober}.

In continuity with the above, the case of general symmetric finite-range lattice walks is investigated
in the present work, with the following setup.
Section~\ref{key} presents a self-contained derivation
of explicit expressions for all fundamental quantities of interest,
including the distribution of the first positive position of the walker (section~\ref{hfull})
and its moments (section~\ref{hmoments}).
In section~\ref{first} we examine the simplest examples of walks whose range is either 1 or~2.
The case of a discrete uniform step distribution with an arbitrary range~$K$
is then studied in section~\ref{uniform}.
A special emphasis is put on the regime where the range becomes large.
The asymptotic expressions of various quantities are compared with their counterparts
for continuous random walks with a symmetric uniform step length distribution.
Section~\ref{disc} contains a discussion
in which analogies and differences between continuous and discrete walks are highlighted,
and the findings of the present paper are set in perspective with those of earlier works~\cite{I,II}.
Finally, two appendices are devoted respectively to the detailed derivations of selected key results
and to integer renewal processes.

\section{Fundamental observables}
\label{key}

In this section we present a self-contained derivation of general results concerning the fundamental quantities,
valid for arbitrary step distributions of the form~(\ref{rhodef}).

\subsection{Preliminaries}

The generating series of the step distribution,
\beq
\hat\rho(z)=\mean{z^\eta}=\sum_k\rho_k z^k=\frac{\rho_K}{z^K}+\cdots+\rho_K z^K,
\label{hatrho}
\eeq
is a self-reciprocal (or palindromic\footnote{The term palindromic refers to the fact that the sequence
of coefficients reads the same forwards and backwards,
just like a palindrome word such as~\emph{radar}.}) Laurent polynomial in $z$,
obeying $\hat\rho(1/z)=\hat\rho(z)$.
The factorisation of the difference
\beq
\phi(z)=1-\hat\rho(z)
\label{phidef}
\eeq
over its complex zeros plays a central role hereafter.
We have
\beq
\phi(z)=-D(z-1)^2+\cdots
\label{zone}
\eeq
as $z\to1$, where
\beq
D=\frac{\mean{\eta^2}}{2}=\frac12\sum_kk^2\rho_k
\eeq
is the diffusion coefficient of the walk.
Besides the above double zero at $z=1$,
$\phi(z)$ has no other zero on the unit circle.
Setting $z=\e^{\ii\alpha}$, we have indeed
\beq
\phi(\e^{\ii\alpha})=\sum_k\rho_k(1-\cos k\alpha)\ge2\rho_1(1-\cos\alpha).
\eeq
The factorisation of the Laurent polynomial $\phi(z)$ therefore reads
\beqa
\phi(z)
&=&-\frac{\rho_K}{z^K}(z-1)^2\prod_a(z+z_a)(z+1/z_a)
\label{phiprod1}
\\
&=&\rho_K(1-z)(1-1/z)\prod_a\frac{(z_a+z)(z_a+1/z)}{z_a}.
\label{phiprod2}
\eeqa
The second form~(\ref{phiprod2}) makes the invariance under the change of $z$ into $1/z$ manifest.
Here and throughout the following, the label $a$ runs over $1,\dots,K-1$.
The $K-1$ complex numbers $z_a$ are inside the unit circle,
whereas their inverses $1/z_a$ are outside the unit circle.
The zeros of $\phi(z)$ are therefore 1 (double), as well as $-z_a$ and $-1/z_a$.
Henceforth, for the sake of brevity, we simply refer to the $z_a$ as being `the zeros'.
They come in complex conjugate pairs.
For $K$ even, there is also a real zero such that $0<z_a<1<1/z_a$.
In generic circumstances, as the range $K$ is increased,
the zeros come closer to the unit circle and become more evenly spread.
Plots of these zeros will be shown in section~\ref{uniform}
(see figure~\ref{zerosplots}) for discrete uniform step distributions.

The leitmotiv of the present approach is that all quantities of interest
can be expressed as symmetric functions of the complex zeros $z_a$.
A first example,
\beq
D=\rho_K\prod_a(1+z_a)(1+1/z_a),
\label{dprod}
\eeq
is obtained by comparing~(\ref{zone}) and~(\ref{phiprod2}).

\subsection{Enhancement factor}

The enhancement factor of the walk is defined by
\beq
\E=\exp\biggl(\frac12\sum_{n\ge1}\frac{1}{n}\,\prob(x_n=0)\biggr).
\label{edef}
\eeq
In the probabilistic literature,
and especially in the classic books~\cite{feller2,spitzerbook},
$1/\E^2$ is denoted by $c$, whereas $D$ is denoted by $\sigma^2/2$.
For a random walk with a continuous step distribution, $\prob(x_n=0)$ vanishes, and so $\E=1$.
If the step distribution is discrete, like e.g.~for the lattice walks considered here,
$\prob(x_n=0)$ may be non-zero, at least for some $n$, and so $\E\ge1$.
The enhancement factor $\E$ therefore measures to what extent a given lattice walk
is different from the class of continuous ones.
This quantity enters the expressions of many observables.
To take a characteristic example~\cite{mms,glmax},
we recall that the mean maximum of a long random walk of $n$ steps takes the universal asymptotic form
\beq
\mean{M_n}\approx2\sqrt\frac{Dn}{\pi},
\label{masy}
\eeq
depending only on the diffusion coefficient of the walk,
whereas the mean number of records, or `ladder heights', of the same walk grows as
\beq
\mean{R_n}\approx\frac{2}{\E}\sqrt\frac{n}{\pi}.
\label{rasy}
\eeq
We have furthermore~\cite{mms,glmax}
\beq
M_n\approx\mean{H}R_n.
\eeq
The ratio between the estimates~(\ref{masy}) and~(\ref{rasy}) therefore reads $\mean{H}=\E\sqrt{D}$.
This result (see~\cite{spitzerbook,mms,glmax,spitzer2,glsur}) is recovered in~(\ref{have}).

The enhancement factor can be evaluated as follows.
From the definition~(\ref{xdef}) of the walk, we have $\mean{z^{x_n}}=\hat\rho(z)^n$, and therefore
\beq
\prob(x_n=0)=\oint\frac{\dd z}{2\pi\ii z}\,\hat\rho(z)^n,
\eeq
where the integration contour is the unit circle.
Equation~(\ref{edef}) can thus be recast as
\beq
\E=\exp\biggl(-\frac12\oint\frac{\dd z}{2\pi\ii z}\ln(1-\hat\rho(z))\biggr).
\label{eoint}
\eeq
The integral above can be evaluated using~(\ref{phiprod2}), together with the identity
\beq
\oint\frac{\dd z}{2\pi\ii z}\ln((y+z)(y+1/z))=\int_0^{2\pi}\frac{\dd\alpha}{2\pi}\ln(1+2y\cos\alpha+y^2)=0
\eeq
valid for complex $y$ with $\abs{y}\le1$.
We thus obtain
\beq
\E=\Bigl(\frac{1}{\rho_K}\prod_a z_a\Bigr)^{1/2}.
\label{eprod}
\eeq
Combining this with~(\ref{dprod}) yields
\beq
\E\sqrt{D}=\prod_a(1+z_a).
\label{ewdprod}
\eeq

Two remarks are in order.
First, a similar but different factorisation scheme for finite-range lattice walks
was proposed recently in a work on the survival probability of random walks and L\'evy flights~\cite{glsur}.
Second, in full generality, `diluting' a step distribution,
i.e., introducing a non-zero probability~$\rho_0$ for the walker to stop at every time step,
alters the separate values of $D$ and $\E$, but neither changes the product~$\E\sqrt{D}$,
nor most quantities to be introduced below, including the distribution of $H$, as should~be.

\subsection{Solution to the homogeneous Wiener--Hopf equation}

Let us begin our study with a first fundamental quantity,
namely the solution to the homogeneous Wiener--Hopf equation
\beq
G_k=\sum_{j\ge0}\rho_{k-j}G_j,
\label{homo}
\eeq
normalised by the boundary condition $G_0=1$.
In analogy with the continuous case~\cite{I},
the preceding equation admits a unique solution, growing linearly with $k$.
In the present setting,
the corresponding generating series,
\beq
\hat G(z)=\sum_{k\ge0}G_k z^k,
\label{hatggdef}
\eeq
reads
\beq
\hat G(z)=\frac{1}{\displaystyle(1-z)^2\prod_a(1+z z_a)}.
\label{ggprod}
\eeq
A detailed derivation of this formula is given in~\ref{app1}.

The behaviour of $G_k$ at large $k$ is of special interest.
Expanding~(\ref{ggprod}) as $z\to1$, using~(\ref{ewdprod}), as well as the identities
\beq
\sum_k z^k=\frac{1}{1-z},\qquad
\sum_k k z^k=\frac{z}{(1-z)^2},
\label{idens}
\eeq
we obtain
\beq
G_k\approx\frac{k+\ell}{\E\sqrt{D}}.
\label{glin}
\eeq
The additive constant $\ell$, known as the extrapolation length, is given by
\beq
\ell=1+\sum_a\frac{z_a}{1+z_a}.
\label{ellsum}
\eeq
The corrections to the asymptotic formula~(\ref{glin}) are exponentially small in $k$.

\subsection{Survival probabilities}

We proceed with the study of the survival probabilities
\beq
g_{n,k}=\prob(x_0\ge0,\ x_1\ge0,\ \dots,\ x_n=k\ge0),
\eeq
defined for $n\ge0$ and $k\ge0$,
and of their temporal sums
\beq
g_k=\sum_{n\ge0}g_{n,k}.
\eeq
Just as in the continuous case~\cite{I},
the survival probabilities $g_{n,k}$, especially $g_k$,
are not only of intrinsic interest but also provide an intermediate step
towards our main objective: the distribution of the first positive position.

Conditioning on the last step of the walk,
and setting $x_{n-1}=j$,
yields the following recursion formula for the survival probability:
\beq
g_{n,k}=\sum_{j\ge0}\rho_{k-j}g_{n-1,j},
\eeq
with initial condition $g_{0,k}=\delta_{k0}$.
As a consequence, the quantities $g_k$ obey
\beq
g_k=\delta_{k0}+\sum_{j\ge0}\rho_{k-j}g_j.
\label{ginhomo}
\eeq
The above equation is an inhomogeneous variant of the Wiener--Hopf equation~(\ref{homo}).
Again in analogy with the continuous case~\cite{I}, it has a unique bounded solution.
The corresponding generating series,
\beq
\hat g(z)=\sum_{k\ge0}g_k z^k,
\label{hatgdef}
\eeq
reads
\beq
\hat g(z)=\frac{\E^2}{\displaystyle(1-z)\prod_a(1+z z_a)}.
\label{gprod}
\eeq
A detailed derivation of this formula is given in~\ref{app2}.
Comparing~(\ref{ggprod}) and~(\ref{gprod}) yields the identity
\beq
\hat g(z)=\E^2(1-z)\hat G(z),
\label{gggrel}
\eeq
implying
\beq
g_k=\E^2(G_k-G_{k-1})\qquad(k\ge1),
\eeq
and especially
\beq
g_0=\E^2.
\label{gzero}
\eeq
At large $k$, $g_k$ approaches exponentially fast the limit
\beq
g_\infty=\frac{\E}{\sqrt{D}}.
\eeq
We also mention the product formula
\beq
\hat g(z)\hat g(1/z)=\frac{\E^2}{1-\hat\rho(z)},
\label{gwh}
\eeq
which is one form of the Wiener--Hopf factorisation.

To close, we make contact with the theory of arbitrary lattice walks,
whose range is infinite in general (see,~e.g.,~\cite{spitzer2}).
To do so, we recast the key result~(\ref{gprod}) as
\beq
\ln\frac{\E^2}{\hat g(z)}=\ln(1-z)+\sum_a\ln(1+z z_a).
\eeq
Noticing that $z=1$ and $z=-z_a$ are the zeros of $\phi(z)$ in the closed unit disk
(see~(\ref{phiprod1})),
we obtain the following contour integral representation for any regular function $F(z)$:
\beq
F(1)+\sum_a F(-z_a)=\oint\frac{\dd y}{2\pi\ii}\,F(y)\left(\frac{\phi'(y)}{\phi(y)}+\frac{K}{y}\right).
\label{Fcontour}
\eeq
The second term inside the parentheses reflects the fact that $\phi(y)$ has a pole of order~$K$ at the origin.
It brings no contribution to the integral whenever $F(0)=0$.
This holds for the case at hand, namely $F(y)=\ln(1-zy)$.
We thus obtain
\beq
\ln\frac{\E^2}{\hat g(z)}=\oint\frac{\dd y}{2\pi\ii}\,\ln(1-zy)\,\frac{\phi'(y)}{\phi(y)}
=z\oint\frac{\dd y}{2\pi\ii}\,\frac{\ln\phi(y)}{1-zy},
\eeq
where the second expression is obtained by integration by parts.
Finally, parametrising the unit circle by $y=\e^{\ii\alpha}$, we get
\beq
\hat g(z)=\E^2\exp\left(z\int_0^{2\pi}\frac{\dd\alpha}{2\pi}
\,\frac{z-\cos\alpha}{1-2z\cos\alpha+z^2}\,\ln(1-\tilde\rho(\alpha))\right),
\label{gcontour}
\eeq
where
\beq
\tilde\rho(\alpha)=\hat\rho(\e^{\ii \alpha}),
\eeq
the Fourier transform of the step distribution, is a real, even, and $2\pi$-periodic function of $\alpha$.
Inserting~(\ref{gcontour}) into~(\ref{gggrel}), we have
\beq
\hat G(z)=\frac{1}{1-z}\exp\left(z\int_0^{2\pi}\frac{\dd\alpha}{2\pi}
\,\frac{z-\cos\alpha}{1-2z\cos\alpha+z^2}\,\ln(1-\tilde\rho(\alpha))\right).
\label{ggcontour}
\eeq

So, for finite-range lattice walks, the key results~(\ref{gprod}) and~(\ref{ggprod}) have been given
the contour integral representation~(\ref{gcontour}) and~(\ref{ggcontour}).
Equivalent forms are given in~\cite{spitzer2}.
The latter representations are discrete analogues of the Pollaczek-Spitzer formulas.
They are far more general than~(\ref{ggprod}) and~(\ref{gprod}),
as they hold for arbitrary symmetric lattice walks,
including those with step distributions falling off as a power law, namely
\beq
\rho_k\approx\frac{c}{\abs{k}^{1+\theta}}.
\eeq
For this class of walks, which includes long-ranged diffusive walks for $\theta>2$
and lattice L\'evy flights for $0<\theta<2$,
the Fourier transform $\tilde\rho(\alpha)$ is well-defined,
whereas the Laurent series (discrete Laplace transform) $\hat\rho(z)$ is ill-defined
whenever $z$ is not along the imaginary axis.

\subsection{Distribution of the first positive position}
\label{hfull}

We now address our main objective, namely the distribution of the first positive position.
Let $N$ denote the discrete time at which the first positive position $H=x_N$ is reached.
The joint distribution of these two random variables is
\beqa
f_{n,k}
&=&\prob(N=n,\ H=k)
\nonumber\\
&=&\prob(x_0\le0,\ x_1\le0,\ \dots,\ x_{n-1}\le0,\ x_n=k),
\eeqa
defined for $n\ge1$ and $k\ge1$.
Our main interest lies in the marginal distribution of the first positive position $H$,
\beq
f_k=\prob(H=k)=\sum_{n\ge1}f_{n,k}.
\eeq

In analogy with the continuous case~\cite{I},
conditioning on the last step of the walk,
and setting $x_{n-1}=-j$,
yields the following expression for the joint probabilities $f_{n,k}$:
\beq
f_{n,k}=\sum_{j\ge0}\rho_{k+j}g_{n-1,j}.
\eeq
As a consequence, the probabilities $f_k$ obey
\beq
f_k=\sum_{j\ge0}\rho_{k+j}g_j.
\label{finhomo}
\eeq
The corresponding generating series,
\beq
\hat f(z)=\mean{z^H}=\sum_{k\ge1}f_k z^k,
\label{hatfdef}
\eeq
reads in the present setting
\beq
\hat f(z)=1-\frac{\E^2}{\hat g(z)}.
\label{fgrel}
\eeq
Using~(\ref{gprod}), this can be written as
\beq
\hat f(z)=1-(1-z)\prod_a(1+z z_a).
\label{fprod}
\eeq
The results~(\ref{fgrel}),~(\ref{fprod}) are derived in detail in~\ref{app3}.
Finally, using~(\ref{gcontour}), we obtain a general expression for $\hat f(z)$,
valid for all lattice walks, namely
\beq
\hat f(z)=1-\exp\left(-z\int_0^{2\pi}\frac{\dd\alpha}{2\pi}
\,\frac{z-\cos\alpha}{1-2z\cos\alpha+z^2}\,\ln(1-\tilde\rho(\alpha))\right).
\label{fcontour}
\eeq

The right-hand side of~(\ref{fprod}) is a polynomial in $z$ of degree $K$, vanishing at $z=0$,
as should be, since the probabilities $f_k=\prob(H=k)$ are nonzero for $k=1,\dots,K$ only.
These probabilities admit explicit expressions in terms
of the elementary symmetric polynomials $S_k$ of the zeros $z_a$,
defined through the expansion
\beq
\prod_a(1+z z_a)=\sum_{k=0}^{K-1}S_k z^k.
\label{skdef}
\eeq
We have
\beq
S_0=1,\quad
S_1=\sum_a z_a,\quad
\dots,\quad
S_{K-1}=\prod_a z_a.
\eeq
The expression~(\ref{fprod}) yields
\beq
f_k=S_{k-1}-S_k\qquad(k=1,\dots,K),
\label{fkres}
\eeq
with the convention $S_K=0$.
We have in particular
\beq
f_1=1-\sum_a z_a,\qquad
f_K=\prod_a z_a=\E^2\rho_K.
\label{fkbdy}
\eeq
The formulas~(\ref{fprod}) and (\ref{fkres}) are the main results of this work.

Using~(\ref{fgrel}), the product identity~(\ref{gwh}) translates to
\beq
(1-\hat f(z))(1-\hat f(1/z))=\E^2(1-\hat\rho(z)).
\label{fwh}
\eeq
This alternative form of the Wiener--Hopf factorisation
yields the following quadratic sum rules for the probabilities~$f_k$:
\beqa
&&\sum_{k=1}^K f_k^2=\E^2(1-\rho_0)-1,
\\
&&\sum_{k=1}^{K-m}f_kf_{k+m}=f_m-\E^2\rho_m\qquad(m=1,\dots,K-1).
\eeqa

\subsection{Moments of the first positive position}
\label{hmoments}

The formula~(\ref{fprod}) also provides a route to evaluate the moments of the first positive position,
\beq
\mean{H^m}=\sum_{k=1}^K k^m f_k.
\eeq
The first moment can be obtained by taking the derivative of~(\ref{fprod}) at $z=1$.
We thus recover the result (see~\cite{spitzerbook,mms,glmax,spitzer2,glsur})
\beq
\mean{H}=\prod_a(1+z_a)=\E\sqrt{D}.
\label{have}
\eeq
In analogy with the continuous case~\cite{I},
the evaluation of higher-order moments can be made systematic by rearranging~(\ref{fprod}) as
\beqa
\hat f(z)
&=&1-(1-z)\E\sqrt{D}\prod_a\frac{1+z z_a}{1+z_a}
\nonumber\\
&=&1-(1-z)\E\sqrt{D}\;\exp\Biggl(\sum_a\ln\frac{1+z z_a}{1+z_a}\Biggr)
\nonumber\\
&=&1-(1-z)\E\sqrt{D}\;\exp\Biggl(\sum_{m\ge1}\frac{c_m}{m!}(z-1)^m\Biggr),
\label{fsolser}
\eeqa
where we have introduced the quantities
\beq
c_m=(-1)^{m-1}(m-1)!\sum_a\biggl(\frac{z_a}{1+z_a}\biggr)^m.
\label{cmres}
\eeq
We have in particular
\beq
c_1=\ell-1,
\eeq
where the extrapolation length $\ell$ is given by~(\ref{ellsum}).

Setting $z=\e^s$ and expanding~(\ref{fsolser}) as a power series in $s$ yields
the following explicit expressions for the first few moments of $H$:
\beqa
\mean{H^2}&=&\E\sqrt{D}\,(2c_1+1)=\E\sqrt{D}\,(2\ell-1),
\nonumber\\
\mean{H^3}&=&\E\sqrt{D}\,(3c_2+3c_1(c_1+2)+1),
\nonumber\\
\mean{H^4}&=&\E\sqrt{D}\,(4c_3+6(2c_1+3)c_2+2c_1(2c_1^2+9c_1+7)+1).
\label{hmoms}
\eeqa
The reduced variance of $H$ reads
\beq
V=\frac{\mean{H^2}}{\mean{H}^2}-1=\frac{2\ell-1}{\E\sqrt{D}}-1.
\label{vres}
\eeq
This quantity is non-negative, hence the inequality $2\ell-1\ge\E\sqrt{D}$.

\subsection{Height renewal process}

In analogy with the continuous case~\cite{I},
the successive upper records $X_1,\ X_2,\ \dots$ of the walker's position form a renewal process,
referred to as the `height renewal process'.
The first record coincides with the first positive position.
More generally,
\beq
X_m=\sum_{j=1}^m H_j,
\eeq
where the integer spatial increments $H_j$ between consecutive record positions
are iid copies of the first positive position $H$,
with distribution $\prob(H=k)=f_k$.
We consistently set $X_0=0$.

For a given lattice position $x\ge0$, there is a unique integer $N_x$ such that
\beq
X_{N_x}\le x<X_{N_x+1}.
\eeq
The random integer $N_x$ represents the number of records up to $x$.

The backward length $B_x$ and the forward (or excess) length $E_x$ are defined by
\beq
B_x=x-X_{N_x},\qquad
E_x=X_{N_x+1}-x.
\eeq
We have thus $B_x\ge0$ whereas $E_x>0$.
These lengths obey
\beq
B_x+E_x=X_{N_x+1}-X_{N_x}=H_{N_x+1}.
\eeq

In the present setting of an integer renewal process,
the distributions of $N_x$, $B_x$ and $E_x$
are less known than their continuous counterparts.
For completeness, a self-contained derivation of these distributions is provided in~\ref{appb}.

It is worth emphasising various relationships between the above variables and results obtained earlier.
The solution $G_k$ to the homogeneous Wiener--Hopf equation admits an appealing interpretation,
in analogy with the continuous case~\cite{I}.
Combining~(\ref{gggrel}) and~(\ref{fgrel}) yields the alternative expression
\beq
\hat G(z)=\frac{1}{(1-z)(1-\hat f(z))}.
\eeq
Comparing the formula above with~(\ref{naveres}),
we obtain
\beq
G_k=1+\mean{N_k},
\label{gnrel}
\eeq
where $\mean{N_k}$ denotes the mean number of records up to position $k$.
Substituting into this relation the expansions~(\ref{glin}) of $G_k$ and~(\ref{navelin}) of $\mean{N_k}$
at large $k$,
one recovers the expressions of the first two moments of $H$, namely
\beq
\mean{H}=\E\sqrt{D},\qquad\mean{H^2}=\E\sqrt{D}\,(2\ell-1),
\eeq
in agreement with~(\ref{have}),~(\ref{hmoms}).

The distribution of the equilibrium backward length,
\beq
B=\lim_{x\to\infty}B_x,
\eeq
or, equivalently, of the equilibrium forward (or excess) length $E=B+1$ (see~(\ref{ebrel})),
is obtained by comparing~(\ref{fsolser}) and~(\ref{lbeq}),~(\ref{leeq}).
We thus obtain
\beq
\mean{z^B}=
\prod_a\frac{1+z z_a}{1+z_a}
=\exp\Biggl(\sum_{m\ge1}\frac{c_m}{m!}(z-1)^m\Biggr).
\label{zbres}
\eeq
Using~(\ref{ewdprod}) and~(\ref{skdef}), we get
\beq
\prob(B=k)=\prob(E=k+1)=\frac{S_k}{\E\sqrt{D}}\qquad(k=0,\dots,K-1).
\eeq
The first moments of $B$ and $E$ are
\beqa
\mean{B}&=&c_1,\qquad\mean{B^2}=c_2+c_1^2+c_1,
\nonumber\\
\mean{E}&=&c_1+1,\qquad\mean{E^2}=c_2+c_1^2+3c_1+1.
\eeqa

From~(\ref{zbres}) it follows that
the coefficients $c_m$, explicitly given in~(\ref{cmres}),
are the factorial cumulants of the equilibrium backward length $B$.
Recall (see,~e.g.,~\cite{KS,JKK})
that the moments $\mu_n$ and cumulants $\gamma_n$ of a real random variable $X$ are defined through
\beq
\mean{\e^{sX}}=\sum_{n\ge0}\frac{\mu_n}{n!}\,s^n=\exp\,\Bigl(\sum_{n\ge1}\frac{\gamma_n}{n!}\,s^n\Bigr),
\eeq
while the factorial moments $m_n$ and factorial cumulants $c_n$
of an integer random variable $N$ are defined through
\beq
\mean{z^N}=\sum_{n\ge0}\frac{m_n}{n!}(z-1)^n=\exp\,\Bigl(\sum_{n\ge1}\frac{c_n}{n!}(z-1)^n\Bigr).
\eeq

\section{First examples: range 1 or 2}
\label{first}

\subsection{Range $K=1$}
\label{keq1}

The simplest lattice random walk has range $K=1$.
This is the so-called simple walk,
where we allow a non-zero probability $\rho_0$ to stop at every time step.
We set
\beq
\rho_{\pm1}=p,\qquad\rho_0=1-2p,
\eeq
with $0<p\le1/2$.
In this case,
\beq
\phi(z)=-\frac{p(z-1)^2}{z}
\eeq
has no non-trivial zeros.
We have
\beq
D=p,\qquad\E=\frac{1}{\sqrt{p}}
\eeq
as well as
\beq
\E\sqrt{D}=1,\qquad\ell=1,\qquad f_1=1,\qquad G_k=k+1.
\label{fgpolya}
\eeq
The latter quantities do not depend on $p$, in line with the remark below~(\ref{ewdprod}).

The P\'olya walk is recovered for $\rho_0=0$, i.e., $p=1/2$.
We have then
\beq
D=\frac12,\qquad\E=\sqrt2.
\eeq

\subsection{Range $K=2$}
\label{keq2}

The next case in increasing order of complexity corresponds to the range $K=2$.
For conciseness, we set $\rho_0=0$, and define
\beq
\rho_{\pm1}=\frac{1-a}{2},\qquad\rho_{\pm2}=\frac{a}{2},
\eeq
with $0\le a<1$.
We have
\beqa
\phi(z)
&=&-\frac{(z-1)^2}{2z^2}(a z^2+(1+a)z+a)
\nonumber\\
&=&-\frac{a(z-1)^2}{2z^2}(z+z_1)(z+1/z_1),
\eeqa
where
\beq
z_1=\frac{1}{2a}\left(1+a-\sqrt{(1-a)(1+3a)}\right)
\eeq
increases monotonically from 0 to 1 as $a$ increases from 0 to 1.
We have also
\beq
D=\frac{1+3a}{2},\qquad\E=\sqrt{\frac{2z_1}{a}}.
\eeq
The enhancement factor takes its largest value, $\E=\sqrt2$, for $a\to0$ and $a\to1$,
and reaches its minimum, $\E=\sqrt{3/2}$, for $a=2/3$, i.e., $z_1=1/2$.
We have as well
\beq
\E\sqrt{D}=1+z_1,\qquad\ell=\frac{1+2z_1}{1+z_1}
\eeq
and
\beq
f_1=1-z_1,\quad f_2=z_1,\qquad G_k=\frac{k+\ell}{1+z_1}+(-1)^k\frac{z_1^{k+2}}{(1+z_1)^2}.
\eeq
In particular,
\beq
\mean{H}=1+z_1,\qquad\mean{H^2}=1+3z_1,
\eeq
so that the reduced variance of $H$ (see~(\ref{vres})),
\beq
V=\frac{z_1(1-z_1)}{(1+z_1)^2},
\eeq
vanishes for $a\to0$ and $a\to1$,
and reaches its maximum, $V=1/8$, for $z_1=1/3$, i.e., $a=3/7$.

\section{Discrete uniform distributions}
\label{uniform}

This section is devoted to the family of lattice walks defined by the symmetric discrete uniform distributions
\beq
\rho_k=\frac{1}{2K+1}\qquad(k=-K,\dots,K),
\label{rhouni}
\eeq
parametrised by the range $K\ge1$.
We have
\beq
D=\frac{K(K+1)}{6}
\label{duni}
\eeq
and
\beq
\hat\rho(z)=\frac{1}{2K+1}\sum_{k=-K}^K z^k=\frac{z^{2K+1}-1}{(2K+1)z^K(z-1)},
\label{rhotuni}
\eeq
so that
\beq
\phi(z)=1-\hat\rho(z)=-\frac{(z-1)^2P_K(z)}{(2K+1)z^K},
\eeq
where
\beqa
P_K(z)
&=&\frac{K(K+1)}{2}z^{K-1}+\sum_{m=0}^{K-2}\frac{(m+1)(m+2)}{2}(z^m+z^{2K-2-m})
\nonumber\\
&=&\prod_a(z+z_a)(z+1/z_a)
\eeqa
(see~(\ref{phiprod1}))
is a self-reciprocal polynomial with degree $2K-2$,
whose coefficients are ascending and descending triangular numbers.

\subsection{The first few values of $K$}

We begin by investigating the first few values of the range $K$.

\begin{itemize}

\item
$K=1$.
In this case, the distribution~(\ref{rhouni}) identifies with that of the simple walk,
studied in section~\ref{keq1}, with $p=1/3$.
We have $D=1/3$, as well as
\beq
\E=\sqrt{3},\qquad\ell=1,
\eeq
and, of course, $f_1=1$, hence $\mean{H}=\mean{H^2}=1$ and $V=0$.

\item
$K=2$.
In this case, the distribution~(\ref{rhouni}) is some `diluted' form of that studied in section~\ref{keq2},
with $a=1/2$.
We have $D=1$ and $P_2(z)=z^2+3z+1$, and therefore
\beq
z_1=\frac{3-\sqrt{5}}{2}=0.381966\dots,
\eeq
so that
\beq
\E=\frac{5-\sqrt{5}}{2}=1.381966\dots,\quad
\ell=\frac{15-\sqrt{5}}{10}=1.276393\dots
\eeq
and
\beq
f_1=\frac{\sqrt{5}-1}{2}=0.618033\dots,\quad
f_2=\frac{3-\sqrt{5}}{2}=0.381966\dots
\eeq
so that
\beqa
\mean{H}=\frac{5-\sqrt{5}}{2}=1.381966\dots,
\nonumber\\
\mean{H^2}=\frac{11-3\sqrt{5}}{2}=2.145898\dots,
\nonumber\\
V=\frac{\sqrt{5}-1}{10}=0.123606\dots
\eeqa

\item
$K=3$.
We have $D=2$ and $P_3(z)=z^4+3z^3+6z^2+3z+1$,
so that there is a pair of complex conjugate zeros,
\beq
z_1=a+\ii b,\qquad z_2=a-\ii b,
\eeq
with
\beqa
a=\frac{(1-\sqrt{7})7^{1/4}}{4\sqrt{2}}+\frac{3}{4}=0.276779\dots,
\nonumber\\
b=\frac{(1+\sqrt{7})7^{1/4}}{4\sqrt{2}}-\frac{\sqrt{7}}{4}=0.386864\dots
\eeqa
We have therefore
\beqa
\E=\frac{7+\sqrt{7}}{2\sqrt{2}}-\frac{7^{3/4}}{2}=1.258529\dots,
\nonumber\\
\ell=2-\frac{1}{\sqrt{2}\,7^{1/4}}=1.565279\dots,
\eeqa
and
\beqa
f_1&=&\frac{(\sqrt{7}-1)7^{1/4}}{2\sqrt{2}}-\frac{1}{2}=0.446441\dots,
\nonumber\\
f_2&=&\frac{1-\sqrt{7}}{2}+\frac{7^{1/4}}{\sqrt{2}}=0.327287\dots,
\nonumber\\
f_3&=&1+\frac{\sqrt{7}}{2}-\frac{(1+\sqrt{7})7^{1/4}}{2\sqrt{2}}=0.226270\dots,
\eeqa
so that
\beqa
&&\mean{H}=\frac{7+\sqrt{7}}{2}-\frac{7^{3/4}}{\sqrt{2}}=1.779829\dots,
\nonumber\\
&&\mean{H^2}=\frac{21+5\sqrt{7}}{2}-\frac{(1+4\sqrt{7})7^{1/4}}{\sqrt{2}}=3.792030\dots,
\nonumber\\
&&V=\frac{1-\sqrt{7}}{4\sqrt{7}}+\frac{2\sqrt{7}-1}{2\sqrt{2}\,7^{3/4}}=0.197057\dots
\eeqa

\end{itemize}

\subsection{Scaling behaviour at large $K$}

We now investigate the regime where the range $K$ of the discrete uniform distribution becomes large.
It is to be expected that the first positive position of the lattice walk reads approximately
\beq
H\approx K\,H_c,
\label{htohc}
\eeq
where $H_c$ is the first positive position of the continuous walk
defined by the continuous uniform step distribution
\beq
\rho_c(x)=\frac{1}{2}\qquad(-1\le x\le +1),
\eeq
whose diffusion coefficient is $D_c=1/6$.

Hereafter we use several notations and results from~\cite{I},
with the addition of a subscript $c$ for quantities pertaining to the continuous case.
The distribution of $H_c$ has a density~$f_c(x)$ on $0\le x\le1$,
which can be determined by means of Wiener--Hopf techniques.
We have in particular $f_c(1)=1/2$, whereas
\beq
f_c(0)=\omega_c=-\frac{1}{\pi}\int_0^\infty\dd q\ln(1-\tilde\rho_c(q)),
\eeq
where
\beq
\tilde\rho_c(q)=\frac{1}{2}\int_{-1}^{+1}\dd x\,\e^{-\ii qx}=\frac{\sin q}{q},
\eeq
hence
\beq
\omega_c=-\frac{1}{\pi}\int_0^\infty\dd q\ln\left(1-\frac{\sin q}{q}\right)=1.601537\dots
\label{omegacres}
\eeq

\begin{figure}[!htbp]
\begin{center}
\includegraphics[angle=0,width=.6\linewidth,clip=true]{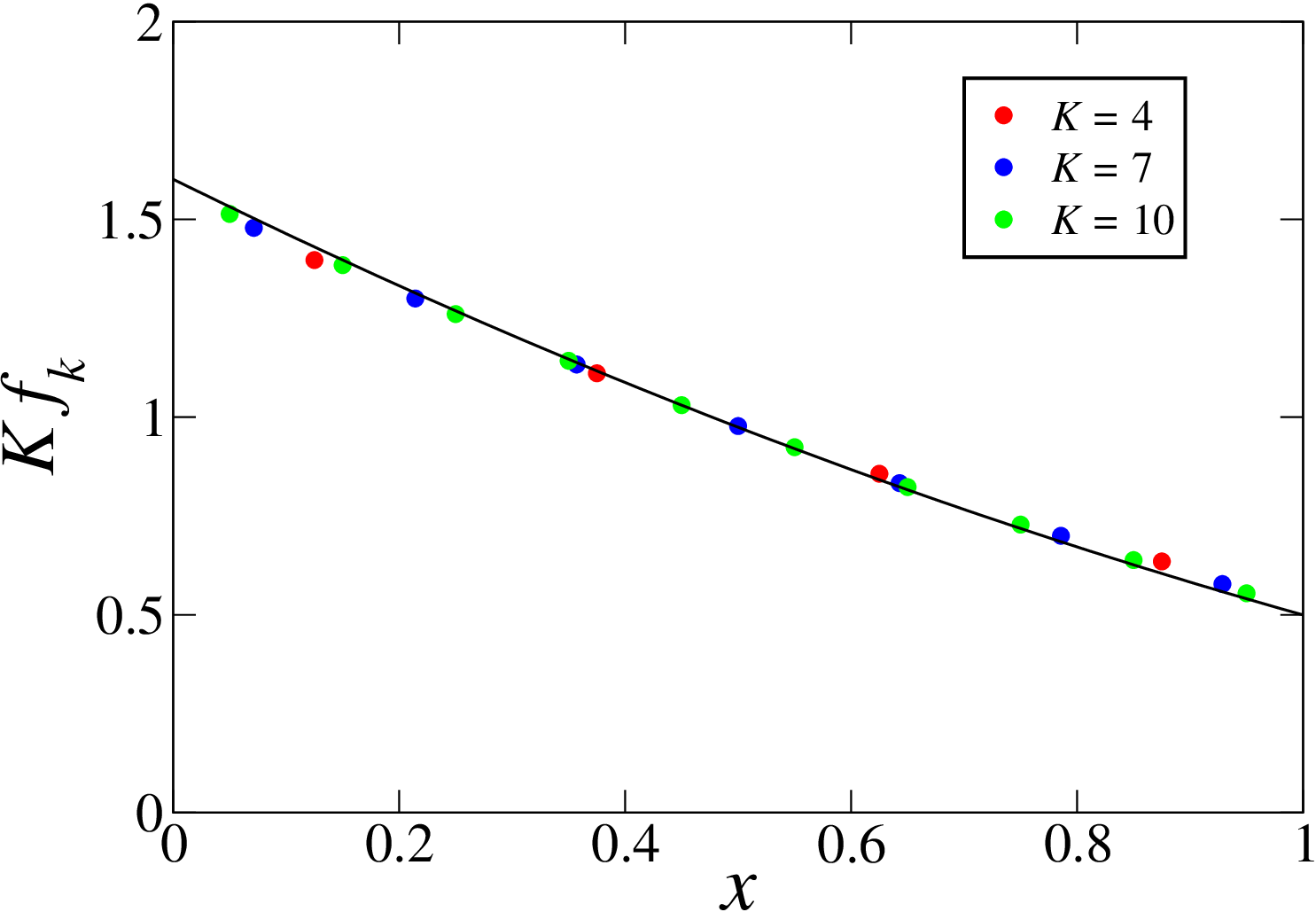}
\caption{
Rescaled probabilities $K f_k$ against $x=(k-1/2)/K$
for discrete uniform step distributions with $K=4$, 7 and 10 (see legend).
Continuous curve: density $f_c(x)$ entering the scaling behaviour~(\ref{fsca}),
obtained by Wiener-Hopf techniques (see~\cite{I}).}
\label{fscaplot}
\end{center}
\end{figure}

In terms of the probabilities $f_k=\prob(H=k)$,
the asymptotic equivalence~(\ref{htohc}) is expected to translate to the scaling behaviour
\beq
f_k\approx\frac{1}{K}\;f_c\left(x=\frac{k-1/2}{K}\right)\qquad(k=1,\dots,K).
\label{fsca}
\eeq
The offset in the numerator has been introduced in order for the $K$ discrete values of $x$
to be distributed symmetrically in the interval $[0,1]$.
Figure~\ref{fscaplot} shows that the convergence of the rescaled probabilities $Kf_k$
to the scaling form~(\ref{fsca}) is quite early,
as a reasonably accurate agreement is already observed for ranges $K$ as small as 4.

This observation will now be corroborated and made quantitative.
Since the diffusion coefficient $D$ is exactly known (see~(\ref{duni})),
we shall, for the time being,
focus on the enhancement factor $\E$ and the extrapolation length $\ell$.
The most direct route to evaluate the behaviour of $\E$ at large $K$ starts from its expression~(\ref{eoint}).
Substituting into it the expression~(\ref{rhotuni}) of $\hat\rho(z)$,
and setting $L=2K+1$ and $z=\e^{\ii \alpha}$, we obtain
\beq
\E=\exp\left(-\frac12\int_0^{2\pi}\frac{\dd \alpha}{2\pi}\ln\biggl(1-\frac{\sin(L\alpha/2)}{L\sin(\alpha/2)}\biggr)\right).
\label{eint}
\eeq
To leading order at large $L$,
changing the integration variable from $\alpha$ to $q=L\alpha/2$ yields
\beq
\E\approx\exp\left(\frac{\omega_c}{2K+1}\right),
\label{easy}
\eeq
where $\omega_c$ is given by~(\ref{omegacres}).
As a consequence of the evenness of the integrand in~(\ref{eint}), both in $L$ and in $\alpha$,
the corrections to~(\ref{easy}) are of relative order $1/L^2$.
For the extrapolation length $\ell$,
the starting point is unavoidably its expression~(\ref{ellsum}),
namely
\beq
\ell=1+\sum_a\frac{z_a}{1+z_a}.
\label{ellsum1}
\eeq
Using again the summation formula~(\ref{Fcontour})
and taking care of the pole at $z=1$, we are left with
\beq
\ell\approx\frac{(2K+1)\ell_c+1}{2},
\label{ellasy}
\eeq
again up to corrections of relative order $1/L^2$.
The extrapolation length of the continuous problem reads
\beq
\ell_c=\frac{1}{\pi}\int_0^\infty\frac{\dd q}{q^2}\ln\frac{D_c\,q^2}{1-\tilde\rho_c(q)},
\eeq
i.e.,
\beq
\ell_c=\frac{1}{\pi}\int_0^\infty\frac{\dd q}{q^2}\ln\frac{q^3}{6(q-\sin q)}=0.297952\dots
\eeq

\begin{figure}[!htbp]
\begin{center}
\includegraphics[angle=0,width=.6\linewidth,clip=true]{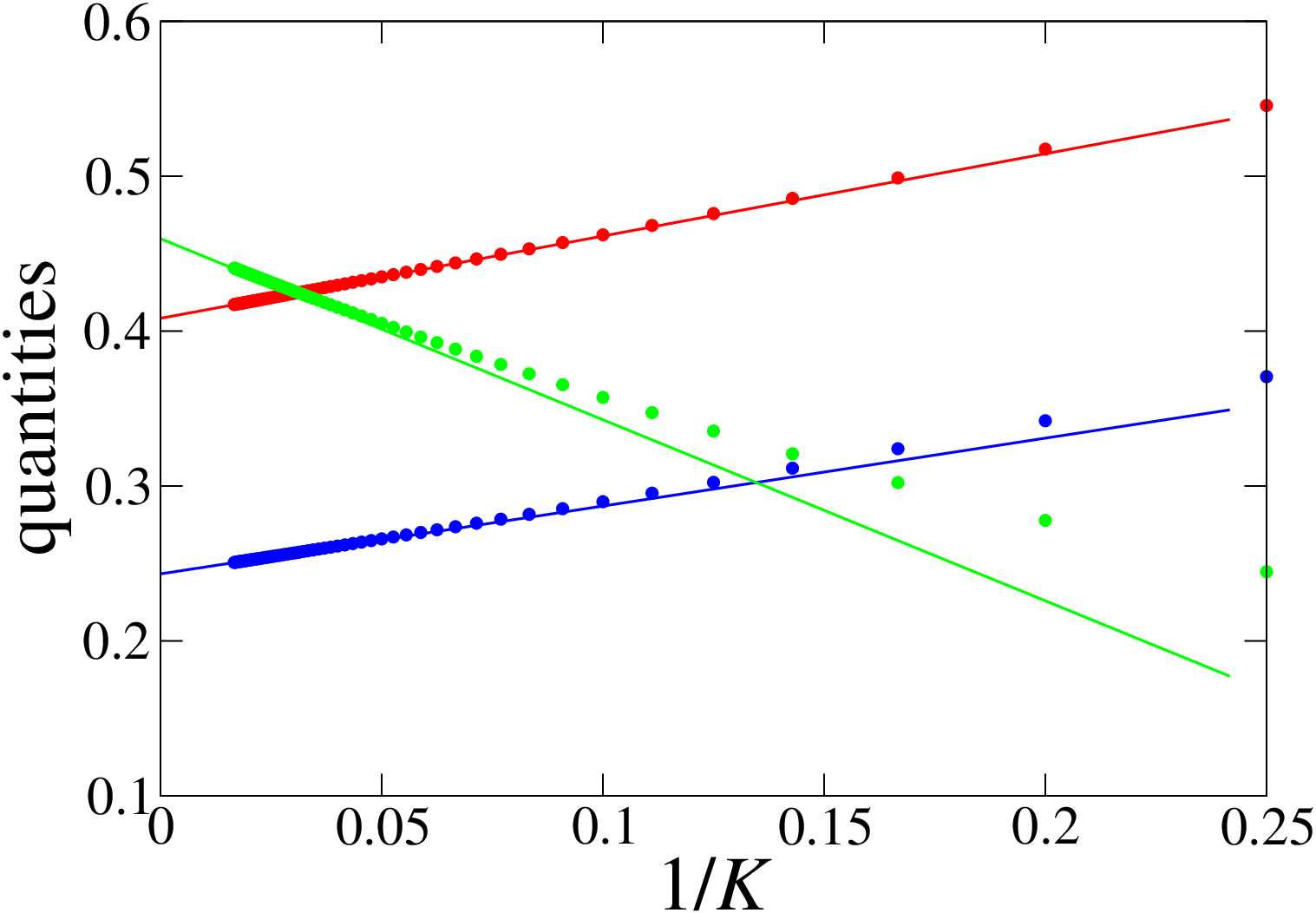}
\caption{
Symbols: plots of $\mean{H}/K$ (red), of $\mean{H^2}/K^2$ (blue) and of $V$ (green)
against $1/K$ for discrete uniform step distributions with $K$ ranging from 4 to 60.
Straight lines with corresponding colours:
expressions~(\ref{momasy}) including linear corrections in $1/K$.}
\label{obsplot}
\end{center}
\end{figure}

Inserting the estimates~(\ref{easy}) and (\ref{ellasy}) into~(\ref{hmoms}),
we obtain the following asymptotic expressions for the first two moments of $H$ and its reduced variance:
\beqa
\mean{H}=\frac{K}{\sqrt{6}}\left(1+\frac{\omega_c+1}{2K}+\cdots\right),
\nonumber\\
\mean{H^2}=\frac{2\ell_c\,K^2}{\sqrt{6}}\left(1+\frac{\omega_c+2}{2K}+\cdots\right),
\nonumber\\
V=2\ell_c\sqrt{6}\left(1-\frac{\omega_c}{2K}+\cdots\right)-1.
\label{momasy}
\eeqa
To leading order at large $K$, these expressions corroborate~(\ref{htohc}), since
\beqa
\mean{H_c}=\frac{1}{\sqrt{6}}=0.408248\dots,
\nonumber\\
\mean{H_c^2}=\frac{2\ell_c}{\sqrt{6}}=0.243276\dots,
\nonumber\\
V_c=2\ell_c\sqrt{6}-1=0.459661\dots
\eeqa
All correction terms in~(\ref{momasy}) involve the quantity $\omega_c$, given in~(\ref{omegacres}).
Figure~\ref{obsplot} demonstrates that the expressions~(\ref{momasy}) provide a good description of the
dependence of the quantities under scrutiny on the range $K$ of the lattice walks.

\begin{figure}[!htbp]
\begin{center}
\includegraphics[angle=0,width=.4\linewidth,clip=true]{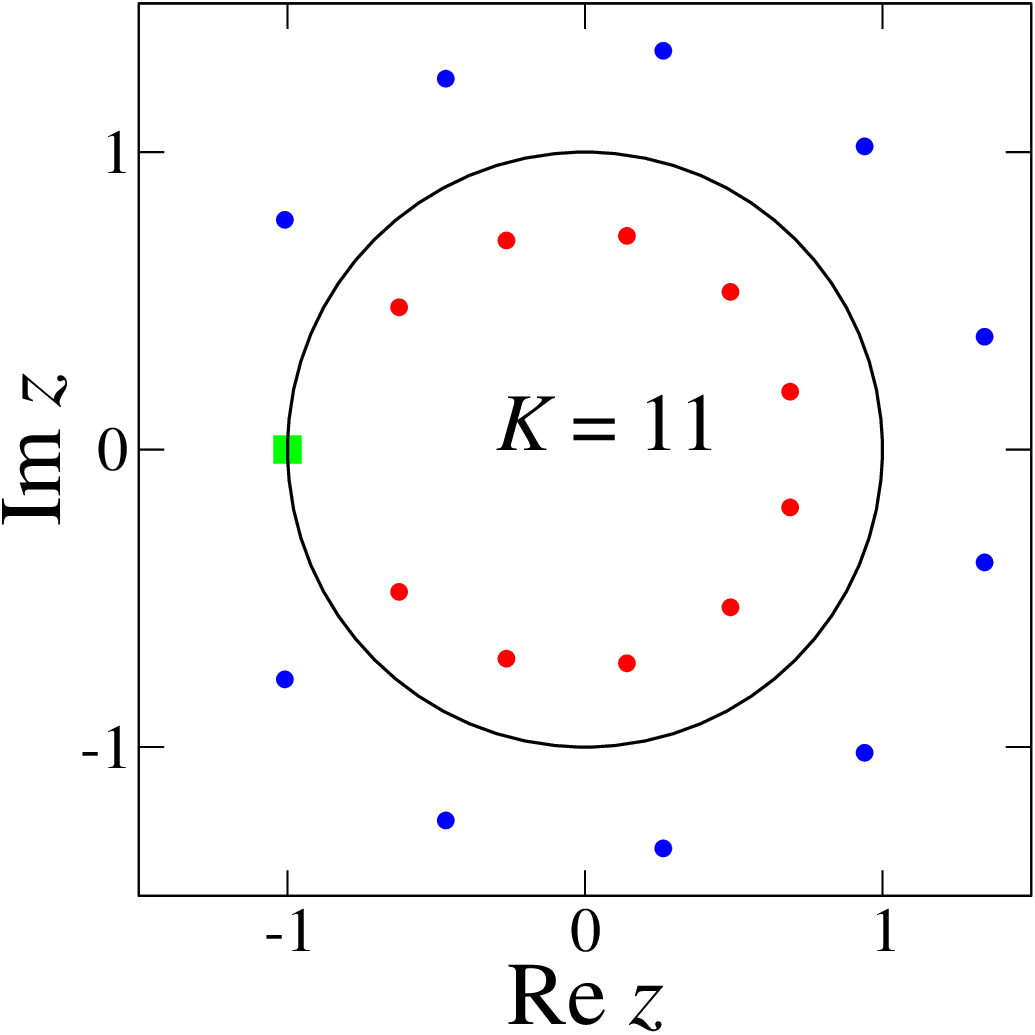}
{\hskip 14pt}
\includegraphics[angle=0,width=.4\linewidth,clip=true]{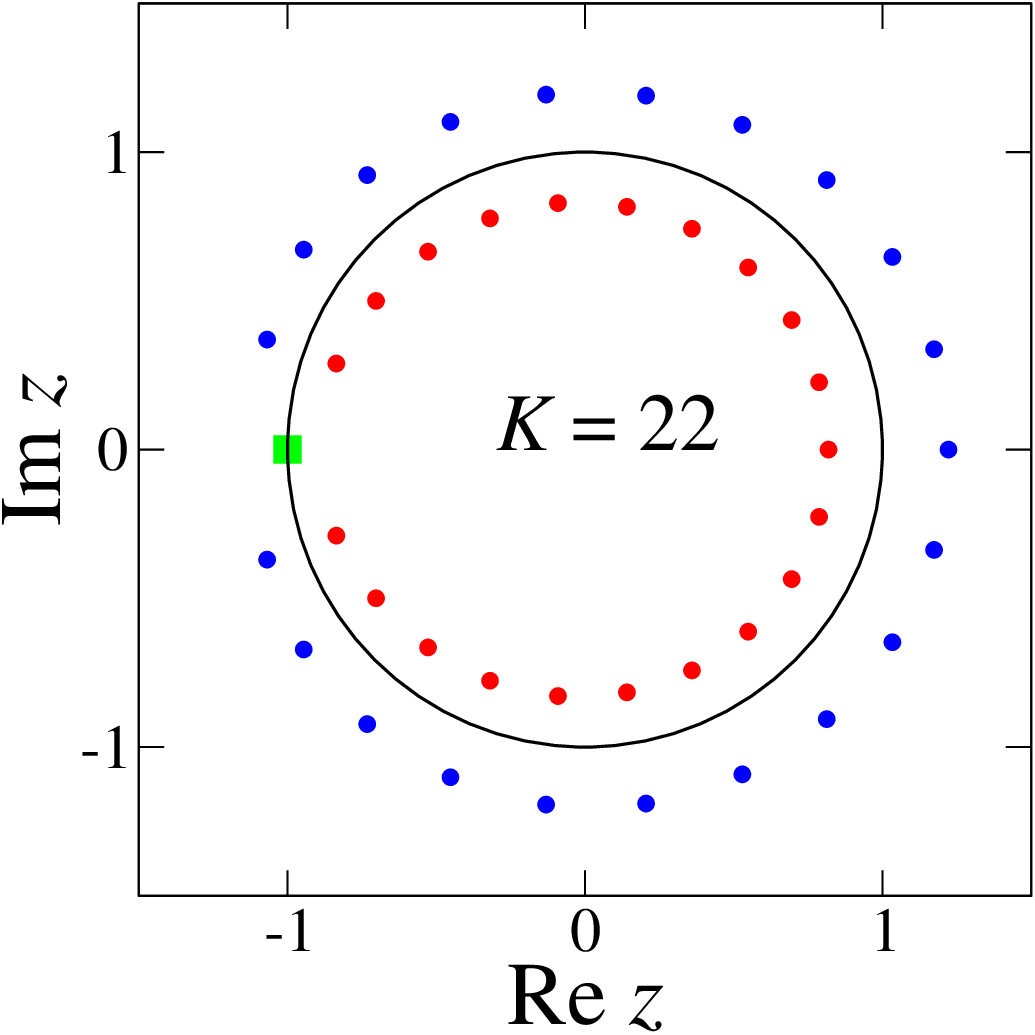}
\caption{
The $K-1$ complex zeros $z_a$ (red symbols) and their reciprocals $1/z_a$ (blue symbols)
in the complex $z$-plane,
for discrete uniform step distributions with range $K$.
Left panel: $K=11$ is odd: there are 10 complex zeros.
Right panel: $K=22$ is even: there are 21 zeros, including a real positive one.
Black: unit circle.
Green square: double diffusive zero at $z=-1$.}
\label{zerosplots}
\end{center}
\end{figure}

To close this analysis of the large-$K$ regime, we examine the behaviour of the zeros~$z_a$,
in terms of which all observables of interest were expressed in section~\ref{key}.
Figure~\ref{zerosplots} confirms the generally expected feature that
zeros approach the unit circle and become more evenly spread as $K$ increases,
leaving some empty space for the double diffusive zero at $z=-1$ (see~(\ref{phiprod1})).

\section{Discussion}
\label{disc}

The present article completes our investigation
of the distribution of the first positive position of a random walker, denoted by $H$.
The case of continuous random walks was analysed in an earlier work~\cite{I},
while the present one is devoted to finite-range lattice walks.
In both papers, only symmetric walks are considered for definiteness,
and the number of steps is taken as a discrete time variable.
Recall that the distribution of~$H$ is the building block
for studying the statistics of all record positions,
or `ladder heights', in one-dimensional random walks,
with these successive positions constituting the `height renewal process'.
In what follows, we highlight a few salient analogies and differences between continuous and lattice walks.

We begin with a feature common to both continuous and lattice walks.
In each setting, there are remarkable classes of step distributions for which
the Wiener-Hopf equations governing the problem can be solved analytically by simple means,
without resorting to the general formalism.
As recalled in the Introduction,
the method consists in factorising the relevant analytic functions over their complex zeros and poles.
In the continuous setting, this approach has long been known:
it applies when the step distribution is a finite superposition of decaying exponentials,
so that its Laplace transform is a rational function.
In the discrete setting, the same method is used extensively here for generic finite-range lattice walks.
The resulting expressions agree with those obtained from the full formalism, as they should.
In particular, the key results~(\ref{ggprod}),~(\ref{gprod}), and~(\ref{fprod})
can be recast as the corresponding discrete Pollaczek-Spitzer
formulas~(\ref{ggcontour}),~(\ref{gcontour}), and~(\ref{fcontour}).

A marked difference between the two settings concerns dimensional analysis.
For continuous walks, it makes sense to measure the step lengths in centimeters, say,
and to apply dimensional analysis throughout.
For discrete walks, such as the lattice walks considered here,
non-trivial dimensionless quantities naturally enter the expressions of observables.
The first and foremost example is the enhancement factor~$\E$.
This difference between the continuous and discrete settings is manifest in the expressions of the moments of $H$.
The expressions~(\ref{hmoms}) derived here for lattice walks involve, besides~$\E$,
several different powers of the quantities $c_k$.
Their continuous analogues, derived in~\cite{I},
only involve the homogeneous polynomials of the highest degree,
in line with the above remark on dimensional analysis.

The last important difference between the two settings
that we wish to address concerns the topmost gap between the two largest positions of a very long walk
(see~\cite{mounaix1,mounaix2}, and~\cite{evsrev} for a review).
For continuous walks, the distribution of the topmost gap, denoted by $Z$,
was studied extensively in an earlier paper~\cite{II}.
That work relied entirely on the identity
\beq
Z=\min(H_1,H_2),
\label{zmin}
\eeq
which states that the gap can be viewed as the smaller of two independent copies of $H$,
associated with the semi-infinite portions of the walk before and after its maximum,
i.e., its largest position.
A key ingredient in deriving~(\ref{zmin}) is the uniqueness of the maximum of the random walk.
This property clearly holds for continuous step distributions, but fails for discrete ones.
More precisely, for a formally infinite discrete walk, such as the lattice walks considered here,
the multiplicity $\nu$ of the maximum, i.e., the number of times it is reached,
is itself a random variable,
whose distribution can be obtained from a formula for the first return time to the origin
of a walker whose position is constrained to stay non-negative at all times (see~\cite[Ch.~XVIII]{feller2}).
We thus arrive at a geometric law,
\beq
\prob(\nu=k)=(1-q)q^{k-1}\qquad(k=1,2,\dots),
\eeq
with parameter
\beq
q=1-\frac{1}{\E^2},
\eeq
where $\E$ is again the enhancement factor.
In particular, the mean multiplicity of the maximum reads
\beq
\mean{\nu}=\E^2.
\eeq
This expression equals unity for continuous walks, reflecting the uniqueness of the maximum.
It exceeds unity for discrete walks.
For instance, we have~$\mean{\nu}=2$ for the simple P\'olya walk.
The non-uniqueness of the maximum invalidates the identity~(\ref{zmin}),
thereby making the study of the topmost gap considerably more difficult in the discrete setting.

\section*{Data availability statement}

Data sharing not applicable to this article.

\section*{Conflict of interest}

The authors declare no conflict of interest.

\section*{Orcid ids}

\noindent
Claude Godr\`eche: https://orcid.org/0000-0002-1833-3490

\noindent Jean-Marc Luck: https://orcid.org/0000-0003-2151-5057

\appendix

\section{Detailed derivations of central formulas}
\label{appa}

\subsection{Derivation of~(\ref{ggprod})}
\label{app1}

\begin{itemize}

\item[1.]
In the definition~(\ref{hatggdef}) of the generating series $\hat G(z)$, i.e.,
\beq
\hat G(z)=\sum_{k\ge0}G_k z^k,
\eeq
which is expected to be analytic for $\abs{z}<1$,
replace $G_k$ by the right-hand side of~(\ref{homo}), obtaining
\beq
\hat G(z)=\sum_{k\ge0}z^k\sum_{j\ge0}\rho_{k-j}G_j.
\eeq

\item[2.]
Express $\rho_{k-j}$ and $G_j$ as the contour integrals
\beq
\rho_{k-j}=\oint\frac{\dd x}{2\pi\ii x^{k-j+1}}\,\hat\rho(x),\qquad
G_j=\oint\frac{\dd y}{2\pi\ii y^{j+1}}\,\hat G(y),
\eeq
obtaining
\beq
\hat G(z)=\sum_{k\ge0}z^k\sum_{j\ge0}\oint\frac{\dd x}{2\pi\ii x^{k-j+1}}\,\hat\rho(x)
\oint\frac{\dd y}{2\pi\ii y^{j+1}}\,\hat G(y).
\eeq

\item[3.]
Perform the geometric sums over $k$ and $j$ to obtain
\beq
\hat G(z)=\oint\frac{\dd x}{2\pi\ii}\,\frac{\hat\rho(x)}{x-z}\oint\frac{\dd y}{2\pi\ii}\,\frac{\hat G(y)}{y-x}
\eeq
for $\abs{z}<\abs{x}<\abs{y}<1$.

\item[4.]
The integral over $y$ is given by the residue of the pole at $y=x$, namely
\beq
\hat G(z)=\oint\frac{\dd x}{2\pi\ii}\,\frac{\hat\rho(x)\hat G(x)}{x-z}
\eeq
for $\abs{z}<\abs{x}<1$.

\item[5.]
Shrink the $x$ contour to $\abs{x}<\abs{z}$, picking the residue of the pole at $x=z$, i.e.,
\beq
\hat G(z)=\hat\rho(z)\hat G(z)+\oint\frac{\dd x}{2\pi\ii}\,\frac{\hat\rho(x)\hat G(x)}{x-z}.
\eeq
Using the definition~(\ref{phidef}) of $\phi(z)$, this reads
\beq
\phi(z)\hat G(z)=\oint\frac{\dd x}{2\pi\ii}\,\frac{\hat\rho(x)\hat G(x)}{x-z}
\label{phigg}
\eeq
for $\abs{x}<\abs{z}<1$.

\item[6.]
As a consequence of the form~(\ref{hatrho}) of $\hat\rho(z)$,
the right-hand side evaluates to a linear combination of the monomials $1/z$ up to $1/z^K$, and therefore
\beq
\phi(z)\hat G(z)=\frac{P_1(z)}{z^K},
\eeq
where $P_1(z)$ is a polynomial of degree $K-1$ that can be determined as follows.
For $z=-z_a$, where $z_a$ is any of the zeros introduced in the factorisation~(\ref{phiprod1}),
$\phi(z)$ vanishes while $\hat G(z)$ is finite, and so $P_1(z)$ vanishes.
We have therefore
\beq
P_1(z)=C\prod_a(z+z_a).
\eeq
The boundary condition $\hat G(0)=G_0=1$ fixes the constant $C$, namely
\beq
C=-\frac{\rho_K}{\displaystyle\prod_a z_a}=-\frac{1}{\E^2}
\eeq
(see~(\ref{eprod})).
We have therefore
\beq
\phi(z)\hat G(z)=-\frac{\rho_K}{z^K}\prod_a(1+z/z_a),
\eeq
hence, using~(\ref{phiprod1}),
\beq
\hat G(z)=\frac{1}{\displaystyle(1-z)^2\prod_a(1+z z_a)},
\eeq
which is~(\ref{ggprod}).

\end{itemize}

\subsection{Derivation of~(\ref{gprod})}
\label{app2}

\begin{itemize}

\item[1.]
In the definition~(\ref{hatgdef}) of the generating series $\hat g(z)$, i.e.,
\beq
\hat g(z)=\sum_{k\ge0}g_k z^k,
\eeq
which is expected to be analytic for $\abs{z}<1$,
replace $g_k$ by the right-hand side of~(\ref{ginhomo}), obtaining
\beq
\hat g(z)=1+\sum_{k\ge0}z^k\sum_{j\ge0}\rho_{k-j}g_j.
\eeq

\item[2.]
Proceed as in~\ref{app1}.
The analogue of~(\ref{phigg}) reads
\beq
\phi(z)\hat g(z)=1+\oint\frac{\dd x}{2\pi\ii}\,\frac{\hat\rho(x)\hat g(x)}{x-z}
\eeq
for $\abs{x}<\abs{z}<1$.

\item[3.]
The integral in the right-hand side again evaluates to a linear combination
of the monomials $1/z$ up to $1/z^K$, and therefore
\beq
\phi(z)\hat g(z)=\frac{P_2(z)}{z^K},
\eeq
where $P_2(z)$ is a polynomial of degree $K$ that can be determined as follows.
First of all, $P_2(z)$ is a monic polynomial: the coefficient of $z^K$ is unity.
For $z=-z_a$, where $z_a$ is any of the zeros introduced in the factorisation~(\ref{phiprod1}),
$\phi(z)$ vanishes while $\hat g(z)$ is finite, and therefore $P_2(z)$ vanishes.
Finally, the $g_k$ being bounded, $g(z)$ only has a simple pole at $z=1$, while $\phi(z)$ has a double pole.
As a consequence, $P_2(z)$ also vanishes for $z=1$.
We have therefore
\beq
P_2(z)=(z-1)\prod_a(z+z_a),
\eeq
hence
\beq
\phi(z)\hat g(z)=\frac{z-1}{z^K}\prod_a(z+z_a),
\eeq
and therefore, using~(\ref{phiprod1}) and~(\ref{eprod}),
\beq
\hat g(z)=\frac{\E^2}{\displaystyle(1-z)\prod_a(1+z z_a)},
\eeq
which is~(\ref{gprod}).

\end{itemize}

\subsection{Derivation of~(\ref{fgrel}) and~(\ref{fprod})}
\label{app3}

\begin{itemize}

\item[1.]
In the definition~(\ref{hatfdef}) of the generating series $\hat f(z)$, i.e.,
\beq
\hat f(z)=\sum_{k\ge1}f_k z^k,
\eeq
which is in fact a polynomial of degree $K$ in $z$,
replace $f_k$ by the right-hand side of~(\ref{finhomo}), obtaining
\beq
\hat f(z)=\sum_{k\ge0}z^k\sum_{j\ge0}\rho_{k+j}g_j.
\eeq

\item[2.]
Express $\rho_{k+j}$ and $g_j$ as the contour integrals
\beq
\rho_{k+j}=\oint\frac{\dd x}{2\pi\ii x^{k+j+1}}\,\hat\rho(x),\qquad
g_j=\oint\frac{\dd y}{2\pi\ii y^{j+1}}\,\hat g(y),
\eeq
obtaining
\beq
\hat f(z)=\sum_{k\ge1}z^k\sum_{j\ge0}\oint\frac{\dd x}{2\pi\ii x^{k+j+1}}\,\hat\rho(x)
\oint\frac{\dd y}{2\pi\ii y^{j+1}}\,\hat g(y).
\eeq

\item[3.]
Perform the geometric sums over $k$ and $j$ to obtain
\beq
\hat f(z)=z\oint\frac{\dd x}{2\pi\ii}\,\frac{\hat\rho(x)}{x-z}\oint\frac{\dd y}{2\pi\ii}\,\frac{\hat g(y)}{xy-1}
\eeq
for $\abs{z}<\abs{x}$ and $\abs{xy}>1$.

\item[4.]
The integral over $y$ is given by the residue of the pole at $y=1/x$, namely
\beq
\hat f(z)=z\oint\frac{\dd x}{2\pi\ii}\,\frac{\hat\rho(x)\hat g(1/x)}{x(x-z)}
\eeq
for $\abs{z}<\abs{x}$.

\item[5.]
Use the identity~(\ref{gwh}) to recast the numerator of the integrand as
\beq
\hat\rho(x)\hat g(1/x)=\hat g(1/x)-\frac{\E^2}{\hat g(x)},
\eeq
so that
\beq
\hat f(z)=z\oint\frac{\dd x}{2\pi\ii}\,\frac{\hat g(1/x)}{x(x-z)}
-z\E^2\oint\frac{\dd x}{2\pi\ii}\,\frac{1}{x(x-z)\hat g(x)}.
\eeq
The first integral vanishes,
as shown by inflating the integration contour to a circle whose radius becomes infinitely large.
The second integral is the sum of the residues of the poles at $x=z$ and $x=0$,
which respectively read $1/(z\hat g(z))$ and $-1/(z\hat g(0))=-1/(z\E^2)$, using~(\ref{gzero}).
Therefore
\beq
\hat f(z)=1-\frac{\E^2}{\hat g(z)},
\eeq
which is~(\ref{fgrel}).
Using~(\ref{gprod}), this can be recast as
\beq
\hat f(z)=1-(1-z)\prod_a(1+z z_a),
\eeq
which is~(\ref{fprod}).

\end{itemize}

\section{Integer renewal processes}
\label{appb}

The notations of this appendix are consistent with those of section~\ref{key}.
We consider an integer renewal process where events sit at positive integer positions:
\beq
X_m=\sum_{j=1}^m H_j.
\eeq
The increments $H_j$ between consecutive events
are iid probabilistic copies of some integer random variable $H$,
distributed according to $\prob(H=k)=f_k$, with generating series
\beq
\hat f(z)=\mean{z^H}=\sum_{k\ge1}f_k z^k.
\eeq
We consistently set $X_0=0$.
For a given position $x\ge0$, there is a unique integer $N_x$ such that
\beq
X_{N_x}\le x<X_{N_x+1}.
\label{nxdef}
\eeq
The random integer $N_x$ represents the number of events up to the given position $x$.
The backward and forward lengths $B_x$ and $E_x$ are defined by
\beq
B_x=x-X_{N_x},\qquad
E_x=X_{N_x+1}-x.
\eeq
We have thus $B_x\ge0$ whereas $E_x>0$.
These lengths obey
\beq
B_x+E_x=X_{N_x+1}-X_{N_x}=H_{N_x+1}.
\eeq

The aim of this appendix is to provide a self-contained derivation of the distributions of $N_x$, $B_x$, and $E_x$.
Special attention will be paid to the equilibrium regime, reached in the limit of a very large observation position $x$.

\subsection{Distribution of $N_x$}

The distribution of the number $N_x$ of events up to position $x$ is conveniently encoded
in the double generating series
\beq
L_N(y,z)=\sum_{x\ge0}z^x\mean{y^{N_x}},
\eeq
which can be determined as follows.
We have
\beqa
L_N(y,z)
&=&\sum_{x\ge0}z^x\sum_{k_1k_2\dots}f_{k_1}f_{k_2}\dots
\nonumber\\
&&\blanc\times\sum_{n\ge0}y^n\;\1(k_1+\cdots+k_n\le x<k_1+\cdots+k_n+k_{n+1})
\nonumber\\
&=&\sum_{n\ge0}y^n\sum_{k_1k_2\dots}f_{k_1}f_{k_2}\dots
\nonumber\\
&&\blanc\times\sum_{x\ge0}z^x\;\1(k_1+\cdots+k_n\le x<k_1+\cdots+k_n+k_{n+1})
\nonumber\\
&=&\sum_{n\ge0}y^n\sum_{k_1k_2\dots}f_{k_1}f_{k_2}\dots(z^{k_1}\dots z^{k_n})(1+z+\cdots+z^{k_{n+1}-1})
\nonumber\\
&=&\sum_{n\ge0}y^n\,\hat f(z)^n\,\frac{1-\hat f(z)}{1-z}
\nonumber\\
&=&\frac{1-\hat f(z)}{(1-z)(1-y\hat f(z))}.
\label{lnres}
\eeqa

The mean number $\mean{N_x}$ of events up to position $x$
is obtained by taking the first derivative of~(\ref{lnres}) at $y=1$, namely
\beq
\sum_{x\ge0}\mean{N_x}z^x=\frac{\hat f(z)}{(1-z)(1-\hat f(z))}.
\label{naveres}
\eeq
The growth of $\mean{N_x}$ at large $x$
can be derived by expanding the formula~(\ref{naveres}) for $z\to1$.
Using
\beq
\hat f(z)=1+\mean{H}(z-1)+(\mean{H^2}-\mean{H})\frac{(z-1)^2}{2}+\cdots,
\eeq
as well as~(\ref{idens}),
we obtain
\beq
\mean{N_x}=\frac{x}{\mean{H}}+\frac{\mean{H^2}+\mean{H}-2\mean{H}^2}{2\mean{H}^2}+\cdots
\label{navelin}
\eeq
The unwritten correction falls off more or less rapidly, depending on the distribution of $H$.
In the situation of interest in this work, where $H$ is bounded by some integer~$K$,
the correction falls off exponentially with $x$.

\subsection{Distribution of $B_x$}

The distribution of the backward length $B_x$ is encoded in the double generating series
\beqa
L_B(y,z)
&=&\sum_{x\ge0}z^x\mean{y^{B_x}}
\nonumber\\
&=&\sum_{x\ge0}z^x\sum_{k_1k_2\dots}f_{k_1}f_{k_2}\dots\sum_{n\ge0}y^{x-(k_1+\cdots+k_n)}
\nonumber\\
&&\blanc\times\1(k_1+\cdots+k_n\le x<k_1+\cdots+k_n+k_{n+1})
\nonumber\\
&=&\sum_{n\ge0}\sum_{k_1k_2\dots}f_{k_1}f_{k_2}\dots\sum_{x\ge0}z^xy^{x-(k_1+\cdots+k_n)}
\nonumber\\
&&\blanc\times\1(k_1+\cdots+k_n\le x<k_1+\cdots+k_n+k_{n+1})
\nonumber\\
&=&\sum_{n\ge0}\sum_{k_1k_2\dots}f_{k_1}f_{k_2}\dots(z^{k_1}\dots z^{k_n})
\nonumber\\
&&\blanc\times(1+yz+\cdots+(yz)^{k_{n+1}-1})
\nonumber\\
&=&\sum_{n\ge0}\hat f(z)^n\,\frac{1-\hat f(yz)}{1-yz}
\nonumber\\
&=&\frac{1-\hat f(yz)}{(1-yz)(1-\hat f(z))}.
\label{lbres}
\eeqa

When the observation position $x$ becomes very large,
the backward length $B_x$ converges to a limit random variable $B$,
the equilibrium backward length,
whose distribution is given by
\beq
\mean{y^B}=\lim_{z\to1}(1-z)L_B(y,z)=\frac{1-\hat f(y)}{(1-y)\mean{H}}.
\label{lbeq}
\eeq

\subsection{Distribution of $E_x$}

The distribution of the forward or excess length $E_x$ is encoded in the double generating series
\beqa
L_E(y,z)
&=&\sum_{x\ge0}z^x\mean{y^{E_x}}
\nonumber\\
&=&\sum_{x\ge0}z^x\sum_{k_1k_2\dots}f_{k_1}f_{k_2}\dots\sum_{n\ge0}y^{k_1+\cdots+k_n+k_{n+1}-x}
\nonumber\\
&&\blanc\times\1(k_1+\cdots+k_n\le x<k_1+\cdots+k_n+k_{n+1})
\nonumber\\
&=&\sum_{n\ge0}\sum_{k_1k_2\dots}f_{k_1}f_{k_2}\dots\sum_{x\ge0}z^xy^{k_1+\cdots+k_n+k_{n+1}-x}
\nonumber\\
&&\blanc\times\1(k_1+\cdots+k_n\le x<k_1+\cdots+k_n+k_{n+1})
\nonumber\\
&=&\sum_{n\ge0}\sum_{k_1k_2\dots}f_{k_1}f_{k_2}\dots(z^{k_1}\dots z^{k_n})
\nonumber\\
&&\blanc\times(y^{k_{n+1}}+y^{k_{n+1}-1}z+\cdots+yz^{k_{n+1}-1})
\nonumber\\
&=&\sum_{n\ge0}\hat f(z)^n\,\frac{y(\hat f(z)-\hat f(y))}{z-y}
\nonumber\\
&=&\frac{y(\hat f(z)-\hat f(y))}{(z-y)(1-\hat f(z))}.
\label{leres}
\eeqa

When the observation position $x$ becomes very large,
the forward length $E_x$ converges to a limit random variable $E$,
the equilibrium forward length,
whose distribution is given by
\beq
\mean{y^E}=\lim_{z\to1}(1-z)L_E(y,z)=\frac{y(1-\hat f(y))}{(1-y)\mean{H}}.
\label{leeq}
\eeq

Comparing~(\ref{lbeq}) and~(\ref{leeq}) yields the identity
\beq
E=B+1.
\label{ebrel}
\eeq
The two lengths coincide for continuous renewal processes at equilibrium.
For integer processes, however, this symmetry is explicitly broken by the definition~(\ref{nxdef}) of $N_x$.

\section*{References}

\bibliography{paper.bib}

\end{document}